\documentclass[a4paper]{article}
\usepackage{spconf}
\usepackage{textcomp}
\usepackage{soul}
\usepackage{graphicx}
\usepackage{amsmath,amssymb}
\usepackage{amsthm}
\usepackage[top=1.75cm,bottom=1.75cm,left=1.8cm,right=1.8cm]{geometry}
\usepackage[english]{babel}
\usepackage[lofdepth,lotdepth]{subfig}
\usepackage{hyperref}
\usepackage{color}
\usepackage[labelfont=bf]{caption}

\usepackage{titlesec}
\titlespacing{\paragraph}{%
  0pt}{%              left margin
  0.2\baselineskip}{% space before (vertical)
  1em}%               space after (vertical)
\titlespacing{\section}{%
  0pt}{%              left margin
  0.3\baselineskip}{% space before (vertical)
  0.5\baselineskip}% space after (vertical)

\usepackage[ruled,vlined]{algorithm2e}
\newcommand{\bpsi}{\boldsymbol{\Psi}}
\newcommand{\bphi}{\boldsymbol{\Phi}}

\newcommand{\norm}[1]{\left\|#1\right\|}

\newcommand{\cn}[1]{\mathbb{C}\mathcal{N}\left(0, #1\right)}

\newcommand{\is}{:=}
\newcommand{\set}[1]{\left[#1\right]}
\newcommand{\sprod}[2]{\langle#1,\,#2\rangle}
\newcommand{\bsprod}[3]{#3\langle#1,\,#2#3\rangle}

\newcommand{\ox}{\Omega_x}
\newcommand{\ov}{\Omega_v}

\newcommand{\IID}{\overset{\text{i.i.d.}}{\sim}}

\definecolor{darkgreen}{rgb}{0.1,0.6,0.2}

\newcommand{\cel}{{\sf c}}

\newcommand{\bs}{\boldsymbol}
\newcommand{\bb}{\mathbb}
\newcommand{\cl}{\mathcal}

\newcommand{\ts}{\textstyle}
\newcommand{\ie}{\emph{i.e.}, }
\newcommand{\eg}{\emph{e.g.}, }

\newcommand{\tiid}{%
    \ifmmode% math mode
        \mathrm{iid}%
    \else%
        iid\xspace%
    \fi%
}

\DeclareMathOperator{\supp}{supp}

\DeclareMathOperator*{\argmax}{arg\,max}

\title{Sparsity-Driven Moving Target Detection\\
in Distributed Multistatic FMCW radars\vspace{-4mm}}
\name{Gilles Monnoyer de Galland de Carni\`eres, Thomas Feuillen, Laurent
Jacques, Luc Vandendorpe\thanks{GM and LJ are funded by the Belgian FNRS.}\vspace{-3mm}}
\address{\ninept ICTEAM/ELEN, UCLouvain, Belgium\vspace{-3mm}}
\newtheorem{theorem}{Theorem}

\begin{document}
\ninept
\maketitle

\begin{abstract}
We investigate the problem of sparse target detection from widely distributed multistatic \textit{Frequency Modulated Continuous Wave} (FMCW) radar systems (using chirp modulation). Unlike previous strategies (\eg developed for FMCW or distributed multistatic radars), we propose a generic framework that scales well in terms of computational complexity for high-resolution space-velocity grid. Our approach assumes that \emph{(i)} the target signal is sparse in a discrete space-velocity domain, hence allowing for non-static target detection, and \emph{(ii)} the resulting multiple baseband radar signals share a common support. By simplifying the representation of the FMCW radar signals, we propose a versatile scheme balancing complexity and detection accuracy. In particular, we design a low-complexity, factorized alternative for the Matching Pursuit algorithm leveraging this simplified model, as well as an iterative methodology to compensate for the errors caused by the model simplifications. Extensive Monte-Carlo simulations of a K-band radar system show that our method achieves a fast estimation of moving target's parameters on dense grids, with controllable accuracy, and reaching state-of-the-art performances compared to previous sparsity-driven approaches.    
\end{abstract}


\section{Introduction}
\label{sec_intro}

Monostatic and bistatic radars with a single transmitter (TX) and a single receiver (RX) only offer the possibility of locating the targets on ellipses and estimating a single component of their velocities. While the addition of antennas to form co-located Multiple Input/Multiple Output (MIMO) radars only enables the estimation of the targets' location vectors, multistatic radars with widely separated TX and RX nodes studied in this paper provide a complete estimation of all target parameters (location and velocity vectors). Moreover, multistatic radars provide spatial diversity by viewing targets from different angles, thereby enhancing the robustness of estimation~\cite{derham2007, derham2010, haimovich2008} and helping target tracking~\cite{gorji2013}, and recognition~\cite{stinco2014}.

The target detection using radars can be formalized as the reconstruction of a sparse vector~\cite{mcclure1997}, \ie with only a few non-zero elements. The support of this vector provides a profile of the targets' locations and velocities. This assumes that the sampled received signal can be decomposed with a few coefficients in a radar sensing matrix (typically a Fourier matrix for FMCW), forming a \emph{dictionary} of waveforms~\cite{hanle1986,herman2009}. Such modeling is instrumental for the application of compressed sensing techniques to radar systems~\cite{candes2008,hadi2015}, and for super resolution radars~\cite{zheng2016, heckel2017}. Recent contributions on multistatic radars also formalized the target detection as the reconstruction of sparse signals~\cite{gogineni2011}. Unlike co-located MIMO systems, multistatic radars do not enable phase-coherent combinations of their signals. While we could treat each received signal independently to obtain estimates of distances and radial speeds, and subsequently combine them (thanks to, \eg triangulation~\cite{handle1987} or trilateration~\cite{ahmad2006}), this paper investigates the joint (across the received signals) modeling and estimation of the target parameters. This leads to a multiple joint reconstruction problem, coupled by a common sparsity pattern~\cite{berger2011}, therefore removes the need of an association step of multiple targets across the radars.

In~\cite{gogineni2011, yu2013, sun2015}, an adaptation of the Matching Pursuit (MP)~\cite{mallat1993}, the Block Matching Pursuit (BMP) suggested by Eldar in~\cite{eldar2010}, is used to exploit the inherent block sparsity of the received signals resulting from the unique sparsity pattern their share. In this context, the 4-D space-velocity domain is sampled into $L$ possible states. However, the value of $L$ can be tremendously large, even at the minimum sampling rate required to fully exploit the potential of the radars. For Pulse Doppler radars (PDRs), properties of the dictionary have been exploited to derive lower complexity strategies in order to overcome this issue~\cite{li2015}.

%corresponding to multistatic FMCW chirp-modulated radar systems 
The low power continuously emitted by FMCW radars makes them more suited for short range applications~\cite{saponara2018}, such as the automotive world~\cite{lin2018, capobianco2018, dokhanchi2018}. However, established dictionaries for FMCW multistatic radars only considered delay estimation in RXs; In other words, only the target localization is considered without velocity estimation~\cite{berger2011}. The first purpose of this paper is therefore to extend this contribution by including the velocity of the targets. Our complete definition of the dictionaries enables us to derive lower complexity algorithms in a factorized fashion. To achieve this goal, we apply relevant simplifications to the complete signals model. Next, a careful analysis of the model mismatches due to the simplifications enable us to provide a bound on the estimation errors of the low complexity algorithm we derived. We then introduce an enhanced algorithm providing an iterative improvement of the target parameters' estimation, rectifying these errors while preserving the low algorithmic complexity order. The algorithms we introduce are evaluated via extensive Monte-Carlo simulations.

%Therefore, the second purpose of our work is the derivation of new strategies to reduce the computation cost while reaching desired grid density using FMCW multistatic radar systems. To achieve this goal, we apply relevant simplifications to the complete signals model we derived and study the losses engendered in performances. Next, we introduce and evaluate an iterative compensation rectifying these losses without increasing the algorithmic complexity order. The mentioned evaluations are performed via extensive Monte-Carlo simulations.

%Such formalization is expected to enable novel efficient applications of CS algorithms, and super-resolution algorithms to multistatic FMCW radars in future work.

\paragraph*{Notations and Conventions:} Matrices and vectors are denoted by bold symbols, $[\bs A]_{n}$ or $\bs A_n$ is the $n$-th column of a matrix $\bs A$, $j=\sqrt{-1}$, and $\cel$ is the speed of light. The scalar product between the vectors $\bs{a}$ and $\bs{b}$ reads $\sprod{\bs{a}}{\bs{b}}$. The transpose and conjugate transpose of a matrix $\bs{A}$ are $\bs{A}^\top$ and $\bs{A}^H$, respectively. The modulo operator is ${\rm mod}$, $\circledast$ is the convolution operator, $[N] := \{1,\cdots,N\}$, and $\cn{\sigma^2}$ is the centered complex normal distribution of variance $\sigma^2$.
\vspace{-1mm}

\section{System and signal description}
\vspace{-1.5mm}
\label{sec_model}
%\TF{I would add in the intro explicitly that we don't care about the implementation but only the signal processing}
From any multistatic radar system, we acquire $Q$ distinct signals, regardless of the practical implementation. Each received signal originates from a bistatic pair of nodes, described by a TX-RX pair such that RX receives, converts adequately into baseband equivalent, and samples the echo resulting from the signal transmitted by TX and reflected by the targets. For instance, a time division multiplexing between transmitters using a shared frequency band can achieve up to $Q = N_T\times N_R$, if composed of $N_T$ TX nodes and $N_R$ RX nodes. A FMCW radar transmits a signal whose waveform is described by \vspace{-1.5mm}
\begin{equation}
    s_T(t) = e^{j2\pi\int_0^t f_c(t')\mathrm{d}t'},\vspace{-1.5mm}
    \label{eq_mod_sT}
\end{equation}
where $f_c(t)$ is the instantaneous carrier frequency at instant $t$. For the chirp modulated radars studied here,\vspace{-1.5mm}
\begin{equation}
    \ts f_c(t) = f_0 + B (\frac{t}{T}\!\!\!\!\mod 1),\vspace{-1.5mm}
    \label{eq_mod_fc}
\end{equation}
where $f_0$ is the lower frequency, $B$ is the bandwidth of the transmitted signal and $T$ is the ramp duration. %. $\left(\frac{t}{T} \mod 1\right)$ describes a periodic saw-tooth function
%\footnote{This can be easily generalized to the 3-D space.\TF{I don't find this comment to be relevant}}
%\TF{ $\cl X \is \{\bs{x}_k\}_{k=1}^K$, subscript notation not introduced}
\paragraph*{a) Complete radar Model:} We model a target as a point scatterer moving in a 2-D plane. Given $K$ targets located in $\cl X \is \{\bs{x}_k\}_1^K$ and moving with constant velocities $\cl V \is \{\bs{v}_k\}_1^K$, the transmission of the $q$-th signal ($q\in [Q]$) acquired from one of the $Q$ bistatic antenna pairs follows the following channel model between the transmitted signal $s^q_T(t)$ and the received signal $s^q_R(t)$,\vspace{-1.5mm}
\begin{equation}
    \ts s^q_R(t) = s^q_T(t) \circledast\big[\sum_{k\in\set{K}}\alpha^q_k\delta\big(t-\tau^q_k(t)\big)\big] + \varepsilon(t),\vspace{-1.5mm}
    \label{eq_mod_channel}
\end{equation}
%\IID \cn{1}$, $q\in [Q], k\in [K]
where $\varepsilon(t)$ is an Additive White Gaussian Noise (AWGN), $\alpha^q_k$ are the scattering coefficients which model the effects occurring in the wave reflection process, including the unknown Radar Cross Sections (RCS). For the sake of simplicity, we neglect in our model the effect of clutter and direct cross talk between TXs and RXs antennas. Their study is postponed to a future work. For all $q \in [Q]$, $\tau^q (t)$ is a \textit{Delay-Doppler} term associated to the $q$-th bistatic pair and defined by\vspace{-1mm}
\begin{equation}
\ts \tau^q_k(t) = \frac{1}{\cel}(r^q(\bs{x}_k) + v^q(\bs{x}_k,\bs{v}_k) t),\vspace{-1mm}
\label{eq_mod_taudef}    
\end{equation}
where $r^q(\bs{x})$ is the bistatic range and $v^q(\bs{x},\bs{v})$ is the bistatic speed of the $q$-th bistatic pair, respectively described by\vspace{-1mm}
\begin{align}
    r^q(\bs{x}) &= \norm{\bs{x}_{\rm t}^q-\bs{x}}_2 + \norm{\bs{x}_{\rm r}^q-\bs{x}}_2,
    \label{eq_mor_bidistdef}\\
    v^q(\bs{x}, \bs{v}) &= \ts \bsprod{\frac{\bs{x}_{\rm t}^q-\bs{x}}{\norm{\bs{x}_{\rm t}^q-\bs{x}}_2} + \frac{\bs{x}_{\rm r}^q-\bs{x}}{\norm{\bs{x}_{\rm r}^q-\bs{x}}_2}}{ \bs{v}}{\big},\vspace{-2mm}
    \label{eq_mor_bispeeddef}
\end{align}
%\TF{weird way of saying it, just put coherent, anyway it either is or not}
where $\bs{x}_{\rm t}^q$ and $\bs{x}_{\rm r}^q$, are the respective locations of the TX and RX nodes of the $q$-th bistatic pair. After a coherent demodulation with the carrier $f_c(t)$ of the received signals in~\eqref{eq_mod_channel}, in which we insert~\eqref{eq_mod_fc} and~\eqref{eq_mod_sT}, we get the baseband signal equivalent \vspace{-2mm}
%Inserting~\eqref{eq_mod_fc} and~\eqref{eq_mod_sT} in~\eqref{eq_mod_channel},
\begin{multline}
    e^q_R(t) = e_{\varepsilon}^q(t) + \\
    \ts \sum_{k\in\set{K}}\alpha^q_k e^{-j2\pi \left(f_0\tau^q_k(t) + B\tau^q_k(t)\left(\frac{t}{T}\!\!\!\!\mod 1\right) - \frac{B}{2T}(\tau^q_k(t))^2\right)},\vspace{-2mm}
    \label{eq_mor_baseband}
\end{multline}
where $e^q_{\varepsilon}(t)$ is the baseband equivalent of $\varepsilon(t)$. The signal is sampled at rate $T_s$ with $M_s$ samples acquired per ramp, such that $T = M_sT_s$, $M_r$ ramps are acquired. If $m_s\in\set{M_s}$ is the index of the sample inside a ramp and $m_r\in\set{M_r}$ is the index of the ramp, the sampled received signal for the $q$-th bistatic pair reads\vspace{-1mm} 
$$
y^q[m_s, m_r] \is e^q_R\big((m_r-1)T+m_sT_s\big).
$$
From~\eqref{eq_mod_taudef} and~\eqref{eq_mor_baseband}, $y^q$ can be decomposed as\vspace{-1mm}
\begin{equation}
    %y^q[m_s, m_r] & = \sum_{k\in\set{K}}\alpha^q_k \psi^q_{\bs{x}_k, \bs{v}_k}[m_s] \phi^q_{\bs{x}_k, %\bs{v}_k}[m_r] \theta^q_{\bs{x}_k, \bs{v}_k}[m_s, m_r] + e^q[m_s, m_r],
    %\label{eq_mod_fact_complete}\\ (pas besoin de cette equation)
    y^q[m_s, m_r]  = \ts \sum_{k\in\set{K}}\alpha^q_k d^q_{\bs{x}_k, \bs{v}_k}[m_s, m_r] + e^q_{\varepsilon}[m_s, m_r],\vspace{-1mm}
    \label{eq_mod_complete}
\end{equation}
where we define\vspace{-1mm}
\begin{equation}
    d^q_{\bs{x}, \bs{v}}[m_s, m_r] \is \psi^q_{\bs{x}, \bs{v}}[m_s]\phi^q_{\bs{x}, \bs{v}}[m_r]\theta^q_{\bs{x}, \bs{v}}[m_s, m_r],\vspace{-1mm}
    \label{eq_mod_ddef}
\end{equation}
with \emph{(i)} the \textit{inner signal} $\psi^q_{\bs{x}, \bs{v}}[m_s]$ only depending on the ramp internal index $m_s$ and such that\vspace{-1mm}
    \begin{align}
    \psi^q_{\bs{x}, \bs{v}}[m_s] \is\ & e^{-j2\pi \frac{f_0}{\cel}r^q(\bs{x})}e^{j2\pi\frac{B}{2M_sT_s}\frac{1}{\cel^2}(r^q(\bs{x}))^2}\nonumber\\
    &e^{-j2\pi\frac{1}{\cel}\left(\frac{B}{M_s}r^q(\bs{x}) + f_0T_sv^q(\bs{x}, \bs{v})\right)m_s},
        \label{eq_mod_psidef}
    \end{align}
\emph{(ii)}  the \textit{outer signal} $\phi^q_{\bs{x}, \bs{v}}[m_r]$ defined by,\vspace{-1mm}
    \begin{equation}
        \phi^q_{\bs{x}, \bs{v}}[m_r] \is e^{-j2\pi\frac{f_0}{\cel}T v^q(\bs{x}, \bs{v})m_r},
        \label{eq_mod_phidef}
    \end{equation}
and \emph{(iii)} the \textit{coupling signal} $\theta^q_{\bs{x}, \bs{v}}[m_s, m_r]$, combining both $m_s$ and $m_r$ in\vspace{-2mm}
    \begin{align}
        \theta^q_{\bs{x}, \bs{v}}[m_s, m_r] \is\ & e^{-j2\pi\frac{1}{\cel}\frac{B}{M_s}v^q(\bs{x}, \bs{v})(m_rT+m_sTs)m_s}\nonumber\\
        &e^{j2\pi\frac{B}{M_sT_s}\frac{1}{\cel^2}(r^q(\bs{x})v^q(\bs{x}, \bs{v})(m_rT+m_sT_s))}\nonumber\\
        &e^{j2\pi\frac{B}{2M_sT_s}\frac{1}{\cel^2}(v^q(\bs{x}, \bs{v})(m_rT+m_sT_s))^2}.\vspace{-2mm}
        \label{eq_mod_thetadef}
    \end{align}
Note that we have arbitrarily gathered in $\psi_{\bs x, \bs v}^q[m_s]$ the factors independent of both $m_s$ and $m_r$.
    
\paragraph*{b) Joint-sparse model of radar measurements:} We define a joint acquisition model where all bistatic pairs simultaneously observe a common scene composed of a few moving targets. We assume them to be localized on a given (separable) space-velocity grid with no more than one target per grid cell, \ie $\cl X \subset \Omega_x \is \{\bs \omega^x_{n}\}_{n=1}^{N_x}$ and $\cl V \subset \Omega_v \is \{\bs \omega^v_{\dot n}\}_{\dot n=1}^{N_v}$, for some \emph{location grid} $\ox \subset \bb R^2$ and \emph{velocity grid} $\ov \subset \bb R^2$ made of $N_x$ locations and $N_v$ velocities, respectively. We postpone to a future study the more involved case of \textit{off-grid} targets, as studied in~\cite{abtahi2016} for MIMO radars. From this assumption,~\eqref{eq_mod_complete} can be recast for each $q \in [Q]$ as the decomposition of $\bs{y}^q := (y^q[1,1], \cdots, y^q[M_s,M_r])^\top$ in the dictionary $\bs{D}^q$ of the $q$-th bistatic pair, where\vspace{-1mm}
$$
\bs{D}^q \is [\bs d^q_{1,1}, \bs d^q_{1,2}, \cdots, \bs d^q_{N_x,N_v}] \in\bb C^{M_sM_r\times N_xN_v}.\vspace{-1mm}
$$ 
This dictionary is composed of the $N_xN_v$ atoms defined by\vspace{-1mm} 
$$
\bs d^q_{n,\dot n} \is \big(d^q_{\bs \omega^x_n,\bs \omega^v_{\dot n}}[1,1],d^q_{\bs \omega^x_n,\bs \omega^v_{\dot n}}[1,2],\cdots,d^q_{\bs \omega^x_n,\bs \omega^v_{\dot n}}[M_s,M_r]\big)^\top,\vspace{-1mm}
$$
where $d^q_{n,\dot n}$ is the $q$-th dictionary's atom corresponding to the $n$-th location and $\dot n$-th velocity (with $n \in [N_x]$ and $\dot n \in [N_v]$,). 

We can, then, turn~\eqref{eq_mod_complete} into the model\vspace{-1mm} 
\begin{equation}
    \bs{y}^q = \bs{D}^q\bs{s}^q + \bs{\varepsilon}^q \in\bb C^{M_sM_r}\vspace{-1mm},
    \label{eq_mod_multifull}
\end{equation}
where $\bs{\varepsilon}^q$ is the AWGN of the $q$-th received signal. The $K$-sparse \emph{target vector} $\bs{s}^q := (s^q[1, 1], \cdots, s^q[N_x, N_v])^\top$ is defined from \vspace{-1.5mm}
\begin{equation}
    s^q[n, {\dot n}] = \begin{cases}
    \alpha_k^q & \text{if $(\bs x_k,\bs v_k) = (\bs \omega^x_n, \bs \omega^v_{\dot n})$}\\
    0 & \text{otherwise.}\vspace{-1.5mm}
    \end{cases}
    \label{eq_mod_sqdef}
\end{equation}
The $Q$ vectors $\{\bs{s}^q\}_{q=1}^Q$ thus share a common support, \ie $\supp{\bs s^{q}} = \cl S$ for all $q \in[Q]$, with $\cl S \subset [N_x N_v]$ and $|S|\leq K$.

\paragraph*{c) Simplified radar model:} While the structure of the dictionaries in~\eqref{eq_mod_multifull} is rather intricate, a simplified joint sparse model can be derived from a few simplifications made on the signals and system properties: \emph{(S1)} the transmitted signal is narrowband, \ie $B \ll f_0$; \emph{(S2)} for all $k\in[K], q\in[Q]$, we have $\tau^q_k<T_s$ such that the echo of transmitted a ramp is acquired when its TX is still emitting the same ramp, hence $r^q(\bs{x}_k)<cT_s$; \emph{(S3)} $\ov$ is chosen \emph{unambiguously}, such that the sampling frequency $1/T$ between ramps which samples $\phi^q_{\bs{x}, \bs{v}}[m_r]$ fulfills the Shannon-Nyquist sampling theorem. This means that $\left|v(\bs{x},\bs{v})\right|\leq 2 \|\bs v\|_2 \leq\frac{\cel}{2f_0T} \hspace{1mm}\forall \bs v \in \ov$. These three simplifications lead to the conclusion that the coupling signal in~\eqref{eq_mod_thetadef} respects $\theta_{\bs{x}, \bs{v}}[m_s, m_r] \simeq 1$. Moreover, \emph{(S3)} ensures that the velocity has small impact on the inner signal and\vspace{-1.5mm}
\begin{equation} 
\psi^q_{\bs{x}, \bs{v}}[m_s] \simeq \psi^q_{\bs{x}}[m_s] \is \psi^q_{\bs{x}, \bs{0}}[m_s]\vspace{-1.5mm}
\label{eq_mod_psidefsimple}
\end{equation}
In practice this approximation amounts to neglecting the Doppler effect within each ramp and thus only considering it between the consecutive ramps. It is equivalent in the context of \eqref{eq_mod_complete} to set $\bs v=0$ in $\psi^q_{\bs{x}, \bs{v}}$. This simplification is known to generate distorsions, as studied in~\cite{liu2010, bao2014, feuillen2016}, but often enables the use of 2D-Discrete Fourier Transforms (DFT) for monostatic FMCW radars. However, the shared grids used to establish the joint model of the multistatic radar signals preclude the use of 2D-DFT. Still, the simplification in \eqref{eq_mod_psidefsimple} enables an analog factorization of the model. %\TF{it should be acknowledged that this is already what is being done when you do range doppler processing on FMCW radar with 2D FFT, otherwise we will sound naive in thinkg that this is new }

%or to set $\bs v=0$ in $\psi^q_{\bs{x}, \bs{v}}$, and extends to the multistatic scenario of the range-Doppler coupling simplification, enabling the use of 2D-IDFTs for FMCW monostatic radars while generating distortions studied in~\cite{liu2010, bao2014, feuillen2016}.%\TF{it should be acknowledged that this is already what is being done when you do range doppler processing on FMCW radar with 2D FFT, otherwise we will sound naive in thinkg that this is new }

Together, these simplifications bring this simplified multistatic radar model: \vspace{-2mm}
\begin{multline}
\hspace{-2mm}y^q[m_s, m_r] \simeq \hspace{-1mm}\sum_{k\in\set{K}} \alpha^q_k \psi^q_{\bs{x}_k}[m_s] \phi^q_{\bs{x}_k, \bs{v}_k}[m_r] + e^q_{\varepsilon}[m_s, m_r].\vspace{-3mm}
    \label{eq_mod_multisimplified}
\end{multline}
Interestingly, we can recast~\eqref{eq_mod_multisimplified} as 
%Thanks to the partial decoupling offered by the simplifications ($\bs{x}$ is exclusively linked to $m_s$ in~\eqref{eq_mod_psidefsimple} \LJ{$\to$ This is obscure, please develop}), we can establish 
%\LJ{sparse matrix does not mean anything; rather speak of target matrix everywhere, or something else, which happens to be sparse according to our priors.}
a factorized joint model for multistatic FMCW radars by defining the $q$-th measurement matrix $\bs{Y}^q \in \bb C^{M_s\times M_r}$ and target matrix $\bs{S}^q \in\bb C^{N_x\times N_v}$ as $(\bs{Y}^q)_{m_s,m_r} := y^q[m_s,m_r]$, and $(\bs{S}^q)_{n,\dot n} := s^q[n,\dot n]$,  
for all $m_s \in [M_s], m_r \in [M_r], n \in [N_x], \dot n \in [N_v], q \in [Q]$. The noise matrices $\bs E^q \in \bb C^{N_x\times N_v}$ are defined similarly from $e^q_\varepsilon$. We also introduce the inner dictionary $\bpsi^q \in\bb C^{M_s\times N_x}$ and outer dictionary $\bphi^q_{n}\in\bb C^{M_r\times N_v}$ such that $(\bpsi^q)_{m_s,n} := \psi_{\bs \omega^x_{n}}[m_s]$, and $(\bphi^q_{n})_{m_r, \dot n} := \phi_{\bs \omega^x_{n}, \omega^v_{\dot n}}[m_r]$, for all $m_s \in [M_s], m_r \in [M_r], n \in N_x, \dot n \in N_v, q \in [Q]$. 

\begin{subequations}
In this setting, the matrix form\vspace{-1.5mm} of~\eqref{eq_mod_multisimplified} is ``factorized" as
\begin{align}
    \bs Y^q&\simeq \bpsi^q \bs{P}^q + \bs{E}^q,\label{eq_mod_simplemat_Y}\\
    \bs P^q&=\big[\bphi_{1}^q [\bs{S}^q{}^\top ]_1, \cdots, \bphi_{N_x}^q [\bs{S}^q{}^\top]_{N_x} \big]^\top.\vspace{-1mm}
    \label{eq_mod_simplemat_P}
\end{align}
Note that, from~\eqref{eq_mod_sqdef}, $\bs S^q$ has no more than $K$ non-zero rows and columns. Therefore, from~\eqref{eq_mod_simplemat_P}, the matrices $\bs P^q$ have less than $K$ non-zero columns, \ie they are \emph{row-sparse}. 
\label{eq_mod_simplemat}
\end{subequations}

\section{Signal recovery algorithms}
\vspace{-1mm}
\label{sec_algo}

The field of computational harmonic analysis provides many algorithms to sparsely decompose a signal in a given dictionary, \eg Basis Pursuit (BP)~\cite{tropp2004}, Matching Pursuit (MP)~\cite{mallat1993}, and MP improvements such as Orthogonal Matching Pursuit (OMP)~\cite{pati1993, foucart2013}. To take advantage of the shared sparsity pattern of the vectors $\bs{s}^q$, $q\in[Q]$ in~\eqref{eq_mod_multifull}, we can use the \emph{Block Matching Pursuit} (BMP)~\cite{eldar2010} that is well suited for multistatic radar applications (see below). As BMP is not adapted to dense grids~\cite{li2015}, we introduce the Factorized Block Matching Pursuit (FBMP), suited for partially factorized models such as~\eqref{eq_mod_multisimplified}, an the Iterative-FBMP (IFBMP) that iteratively corrects the errors committed by FBMP.

\paragraph*{a) Block Matching Pursuit:} %\LJ{There were errors in Alg. 1 that I corrected. Also, we had to say the inputs/outputs.}
Alg.~\ref{alg_BMP_applied} directly formulates BMP to the joint multistatic radars model~\eqref{eq_mod_complete}. BMP proceeds by jointly processing the $Q$ bistatic TX-RX pairs: at each iteration, BMP first selects in~\eqref{eq_alg_bmpind} the indices $\left\{n^*,{\dot n}^*\right\}$ corresponding to the atoms collaboratively maximizing their correlation with the $Q$ residuals $\bs r^{q,(k)}$ (each set to $\bs y^q$ at $k=0$), stores these correlations in $\hat{s}^q$, and finally updates these residuals by removing the contribution of the selected atoms. The computational complexity of BMP is dominated by~\eqref{eq_alg_bmpind} as this step requires us to compute the contributions of all location-velocity indices to pick the largest one. Overall, the complexity scales like $O(K Q M_r M_s N_x N_v)$ at each iteration, which becomes quickly unaffordable for dense grids $\ox$ and $\ov$.

\begin{algorithm}[t]%[H]
%\SetAlgoLined
\SetKwInOut{Input}{Input}\SetKwInOut{Output}{Output}
\SetKwFor{While}{While}{:}{end}

%\Input{Number of targets $K$, \hspace{0.1cm} measurement vectors $\bs{y}^q$.}
%\\ Location Grid $\ox$ of size $N_x$ and Velocity Grid $\ov$ of size $N_v$
%\Output{sparse vectors estimates $\hat{\bs{s}}^q$,  for $q\in[Q]$.}% \\
%Locations estimates $\left\{\hat{\bs{x}}_k : k\in\set{K}\right\}$\\ 
%Velocities estimates $\left\{\hat{\bs{v}}_k : k\in\set{K}\right\}$.} 

\vspace{0.05cm}

\Input{$\{\bs y^q\}_{q=1}^Q$, $\{\bs D^q\}_{q=1}^Q$, $K$.}
\Output{Estimated target signals $\{\hat{s}^{q}\}_{q=1}^Q$.}
\ \\[-3mm]

Initialization: $k=1$; $\forall q\in[Q], \bs{r}^{q,(0)} = \bs{y}^q, \hat{\bs{s}}^{q,(0)} = \bs{0}$.
\ \\[1mm]

\While{$k < K$}{
%\vspace{-4mm}
\ \\[-7mm]
\begin{align}
&\!\!\!\!\!\!\!\!\!\!\{n^*, {\dot n}^*\} = \!\!\!\!\!\!\argmax_{(n,\dot n)\in [N_x]\times[N_v]} \sum_{q\in [Q]} |\sprod{\bs{d}^q_{n,{\dot n}}}{\bs{r}^{q,(k)}}|^2.
        \label{eq_alg_bmpind}
\end{align}
\vspace{-3mm}

$\hat{s}^{q}[n^*, {\dot n}^*] = \hat{s}^{q}[n^*, {\dot n}^*] + \frac{\sprod{\bs{d}^q_{n^*,{\dot n}^*}}{\bs{r}^{q,(k)}}}{M_rM_s}, \hspace{0.1cm}  \forall q\in [Q].$
\vspace{1mm}

$\bs{r}^{q,(k+1)} = \bs{r}^{q,(k)} - \frac{\sprod{\bs{d}^q_{n^*,{\dot n}^*}}{\bs{r}^{q,(k)}}}{M_rM_s}\,\bs{d}^q_{n^*,{\dot n}^*}, \hspace{0.1cm} \forall q\in [Q].$
\vspace{-3mm}

$k \leftarrow k+1$
}
  \caption{\ninept BMP for multistatic radars}
  \label{alg_BMP_applied}
\end{algorithm}

\paragraph*{b) Factorized BMP:} We propose a new algorithm, the Factorized Block Matching Pursuit (FBMP), which reduces the complexity of BMP by leveraging the factorized model~\eqref{eq_mod_simplemat}, \ie by replacing~\eqref{eq_alg_bmpind} in Alg.~\ref{alg_BMP_applied} by an approximate method. Writing $\bs \psi^q_n$ (and $\bs \phi^q_{n,\dot n}$) for the $n$-th (resp. $\dot n$-th) column of $\bs \Psi^q$ (resp. $\bs \Phi_{n}^q$), and given the residuals $\bs R^{q,(k)} = (\bs r^{q,(k)}_1, \cdots, \bs r^{q,(k)}_{M_r}) \in \bb C^{N_x \times M_r}$ initialized to $\bs R^{q,(0)} = \bs Y^q$ for $q\in [Q]$, we perform at each FBMP iteration the following two-step procedure. 
First, inspired from~\eqref{eq_mod_simplemat_Y}, the location index $n^*$ is computed from the inner dictionaries $\bs \Psi^q$, independently of ${\dot n}^*$, in other words $n^*$ is the location index of the inner atoms collaboratively maximizing their ``intra-ramp'' correlation with all the residuals, $n^*$ is computed by \vspace{-1mm} 
\begin{equation}
\ts n^* = \argmax_{n\in\set{N_x}}\sum_{q\in[Q]}\sum_{m_r\in\set{M_r}} |\sprod{\bs{\psi}^q_{n}}{\bs{r}^{q,(k)}_{m_r}}|^2.\vspace{-1mm}
\label{eq_alg_fbmpnx}
\end{equation}
Next, by considering~\eqref{eq_mod_simplemat_P}, we find ${\dot n}^*$ from the outer dictionaries $\bs \Phi_{n^*}^q$ defined on~$n^*$ for each $q \in [Q]$: $\dot n^*$ is the velocity index corresponding to the outer atoms localized on $n^*$ that collaboratively maximize their correlation with $\bs p_{n^*}^q$, \vspace{-1mm}
\begin{equation}
\ts {\dot n}^* = \argmax_{{\dot n}\in\set{M_r}}\sum_{q\in[Q]}\left|\sprod{\bs{\phi}^q_{n^*,{\dot n}}}{\Tilde{\bs{p}}^q_{n^*}}\right|^2.\vspace{-1mm}
\label{eq_alg_fbmpdotn}
\end{equation}
with $\bs p_{n^*}^q \is (\sprod{\bs{\psi}^q_{n^*}}{\bs{r}^{q,(k)}_1}, \cdots, \sprod{\bs{\psi}^q_{n^*}}{\bs{r}^{q,(k)}_{M_r}})^\top$.

The computational complexity of this two-step procedure reduces to $O(K Q (M_rM_sN_x + M_sN_v))$, which is lower than for~\eqref{eq_alg_bmpind} in BMP for dense grids. However, FBMP suffers from the model mismatches induced by the simplifications introduced in Sec.~\ref{sec_model}, resulting in a wrong estimation of the targets' locations. 

In fact, in noiseless condition, neglecting the influence of the coupling signals, and in the context of a single target ($K=1$) with parameters $(\bs x, \bs v)$ where $\bs v$ is known (in an oracle context), the target location would be more accurately estimated by \eqref{eq_alg_fbmpnx} if the atoms of the inner dictionaries $\bpsi^q_{\bs v}$ such that $(\bpsi^q_{\bs v})_{m_s,n} := \psi_{\bs \omega^x_{n},{\bs v}}[m_s]$ were used instead of the atoms $\bs\psi^q_n$ of $\bpsi^q = \bpsi^q_{\bs 0}$. However, Thm.~\ref{th_1} shows that if $\ov$ respects a Shannon-Nyquist criterion, and if the TX and RX nodes are located sufficiently far of the location grid, then for any $q\in[Q]$ and $\bs x\in\ox$, the atom associated with the location $\bs x$ in the dictionary $\bpsi^q_{\bs{v}}$ tends to be proportional to the atom associated to a location $\tilde{\bs x}$ shifted in the direction of $\bs v$ in the dictionary $\bpsi^q$ (assuming $\tilde{\bs x}$ is still in $\ox$). Hence, for a sufficiently dense location grid, the dictionaries $\bpsi^q, q\in[Q]$ approximate the dictionaries $\bpsi^q_{\bs{v}}, q\in[Q]$ with the atoms' indices shifted in a manner that depends on $\bs v$ (see also Sec.~\ref{sec_simu}). 

\begin{theorem}
\label{th_1}
Given some $\lambda > 3$, $\gamma := \frac{f_0M_sT_s}{B}$, and the 2-D positions $\bs x^q_{\rm t}$ and $\bs x^q_{\rm r}$ of the TX and RX antennas for each bistatic pair $q \in [Q]$, respectively, if
\begin{eqnarray}
    \label{eq_alg_vbounded}
    &\!\!\!\!\max_{\bs v \in \Omega_v}|v(\bs{x},\bs{v})|_2\leq 2 \max_{\bs v \in \Omega_v} \|\bs v\|_2 \leq \ts \frac{\cel}{2f_0T},\\
    \label{eq_alg_radoutside}
    &\!\!\!\!\min_{\bs x\in\ox, q\in[Q]} \min\big(\norm{\bs{x}_{\rm t}^q-\bs{x}}_2,\norm{\bs{x}_{\rm r}^q-\bs{x}}_2\big) > \ts \lambda\frac{\cel}{4B},
\end{eqnarray}    
then, there exist a function $A^q(\bs x,\bs v) \in \bb C$ with $|A^q(\bs x,\bs v)|=1$, independent of $m_s$, such that, for all $m_s \in [M_s]$ and $q \in[Q]$, 
\begin{equation}
    \ts \big|\psi^q_{\bs{x},\bs{v}}[m_s] - A^q(\bs x,\bs v) \psi^q_{\tilde{\bs x},\bs{0}}[m_s]\big| = O(\frac{1}{\lambda}),
\end{equation}
where $\tilde{\bs x} := \bs{x}+\gamma\bs{v}$ and $\|\tilde{\bs x} -\bs x\|_{\infty} \leq \frac{\cel}{4\sqrt{2}B}$.\vspace{-2mm}
\end{theorem}

Because we always have $|A^q(\bs x,\bs v)|=1$ the above reasoning implies that, for a highly dense location grid, FBMP tends to estimate $\hat{\bs{x}} = \bs \omega^x_{n} = \tilde{\bs x}$ instead of $\bs x$. Moreover, the bound on the estimation error $\|\tilde{\bs x} -\bs x\|_{\infty}$ provided by Thm.~\ref{th_1} asserts that, for a sufficiently large value of $\lambda$, the estimated location $\hat{\bs{x}}$ is close to the exact target's location $\bs{x}$ compared to the distance between the target and the TX and RX nodes. This implies that the scalar product in \eqref{eq_mor_bispeeddef}, and hence the chosen outer dictionaries for the velocity estimation, $\bphi^q_{n}$, $q\in[Q]$, are only poorly affected by the location estimation error. Therefore, $\hat{\bs{v}} = \bs \omega^v_{\dot n^*}$ is also close to the exact target's velocity $\bs{v}$ when compared to the initial arbitrary guess $\hat{\bs{v}}=0$.

%the first guess for the location estimation: $\hat{\bs{x}}_{(0)} = \bs \omega^x_{n^0}$ (the closest bin of $\ox$ to $\tilde{\bs x}$) is close to the exact target's location $\bs{x}$ compared to the distance between the target and the TX and RX nodes. This implies that the scalar product in \eqref{eq_mor_bispeeddef}, and hence the chosen outer dictionary for the velocity estimation, $\bphi^q_{n^0}$, is only poorly affected by the location estimation error. Therefore, $\hat{\bs{v}}_{(0)} = \omega^v_{\dot n^0}$ is also close to the exact target's velocity $\bs{v}$ when compared to the initial arbitrary guess $\hat{\bs{v}}_{(-1)}=0$. %Moreover, for dense grids, the linearity in $\tilde{\bs x} := \bs{x}+\gamma\bs{v}$ implies that an improved velocity estimation involves, in the next iteration of eq. \eqref{eq_alg_ifbmpn}, either an improvement or no change in the location estimation, depending on the location grid's density. 

\paragraph*{c) Iterative FBMP (IFBMP):} We can compensate the errors caused by the model mismatch committed by FBMP by iteratively improving the space-velocity indices of each selected atom, and this with only a slight increase in computational complexity. 

IFBMP is similar to FBMP except that at the $k$-th iteration, the selection step is replaced by $N_{\rm it}$ successive improvements of the atom space-velocity parameters according to a sequence of indices $\{(n^i, \dot n^i): 0\leq i \leq N_{\rm it}\}$. A similar strategy was proposed in~\cite{gribonval2001} to improve MP in the selection of atoms in a dictionary made of Gaussian chirps.

Following the notations introduced for FBMP, at initialization, $(n^0, \dot n^0)$ are found from \eqref{eq_alg_fbmpnx} and \eqref{eq_alg_fbmpdotn}. Next, for $i\geq 1$, the $i$-th space-velocity improvement $(n^i, \dot n^i)$ is obtained from 
\begin{align}
\label{eq_alg_ifbmpn}
\ts n^i&= \argmax_{n\in\set{N_x}} \ts \sum_{q\in[Q]}\sum_{m_r\in\set{M_r}} |\sprod{\bs{\psi}^q_{n, \dot n^{i-1}}}{\bs{r}^{q,(k)}_{m_r}}|^2,\\
\label{eq_alg_ifbmpdotn}
\dot n^i& = \argmax_{{\dot n}\in\set{M_r}}\ts \sum_{q\in[Q]} |\sprod{\bs{\phi}^q_{n^i,\dot n}}{\Tilde{\bs{p}}^q_{i}}|^2,\vspace{-2mm}
\end{align}
with $\bs p_{i}^q \is (\sprod{\bs{\psi}^q_{n^i,\dot n^{i-1}}}{\bs{r}^{q,(k)}_1}, \cdots, \sprod{\bs{\psi}^q_{n^i,\dot n^{i-1}}}{\bs{r}^{q,(k)}_{M_r}})^\top$, and 
$\bs \psi^q_{n, \dot n} := (\psi^q_{\bs \omega^x_n, \bs \omega^v_{\dot n}}[1], \cdots, \psi^q_{\bs \omega^x_n, \bs \omega^v_{\dot n}}[M_s])^\top$ for $n \in [N_x]$, $\dot n \in [N_v]$, and $q \in [Q]$. Finally, $(n^*, \dot n^*) = (n^{N_{\rm it}},\dot n^{N_{\rm it}})$ are the returned indices of the $k$-th selected atom of IFBMP.  

%\begin{algorithm}[tb]
%\SetAlgoLined
%\SetKwInOut{Input}{Input}\SetKwInOut{Output}{Output}
%\SetKwFor{While}{While}{:}{end}
%Initialization: $\dot n^*_{(0)} = \amin{{\dot n}}\norm{\omega_{\dot n}^v}_2$, \hspace{2mm} $i:=1$.
% \vspace{-1.5mm}

% \While{$i \leq N_{it}$ and $\{n^*_{(i)}, \dot n^*_{(i)}\}\neq \{n^*_{(i-1)}, \dot n^*_{(i-1)}\}$}{
%     \begin{align*}
%         &\!\!\!\!\!\!n^*_{(i)} = \argmax_{n\in\set{N_x}}\sum_{q\in[Q]}\sum_{m_r\in\set{M_r}}\bigg|\sprod{\bs{\psi}^q_{n,\dot n^*_{(i-1)}}}{\bs{r}^{q,(k)}[m_r]}\bigg|^2.\\
%         %&\!\!\!\!\!\!\!\!\!\!\!\!\Tilde{\bs{p}}_{n^*_{(i)}}^q \is \left(\sprod{\bs{\psi}^q_{n^*_{(i)}}}{\bs{r}^{q,(k)}[1]}, \cdots, \sprod{\bs{\psi}^q_{n^*_{(i)}}}{\bs{r}^{q,(k)}[M_r^q]}\right)^\top.\\
%         &\!\!\!\!\!\!\dot n^*_{(i)} = \argmax_{{\dot n}\in\set{M_r}}\sum_{q\in[Q]}\bigg|\sprod{\bs{\phi}^q_{n^*_{(i)},{\dot n}}}{\Tilde{\bs{p}}^q_{n^*_{(i)}}}\bigg|^2.
%     \end{align*}
%     \vspace{-2mm}
%     $i\leftarrow i+1$
% }
%   \caption{Best matching atoms search in IFBMP}
%   \label{alg_IFBMP}
% \end{algorithm}

The complexity of IFBMP scales like $O(KN_{\rm it}Q\,(M_rM_sN_x + M_sN_v))$. The improvement brought by IFBMP over FBMP, as well as its convergence, strongly depends on the location of the TX and RX antennas and the geometry of the grids $\ox$ and $\ov$; their optimization is postponed to a future study.

\begin{figure*}[tb]
\centering
%% Hauteur des labels, par rapport au bas des figures. Just modifier ici
\newlength{\hlabel}
\setlength{\hlabel}{4.45cm}
%%%%%%%%% Fig 1.(a)
\parbox[b]{0.275\textwidth}{%
\noindent\includegraphics[width=0.275\textwidth]{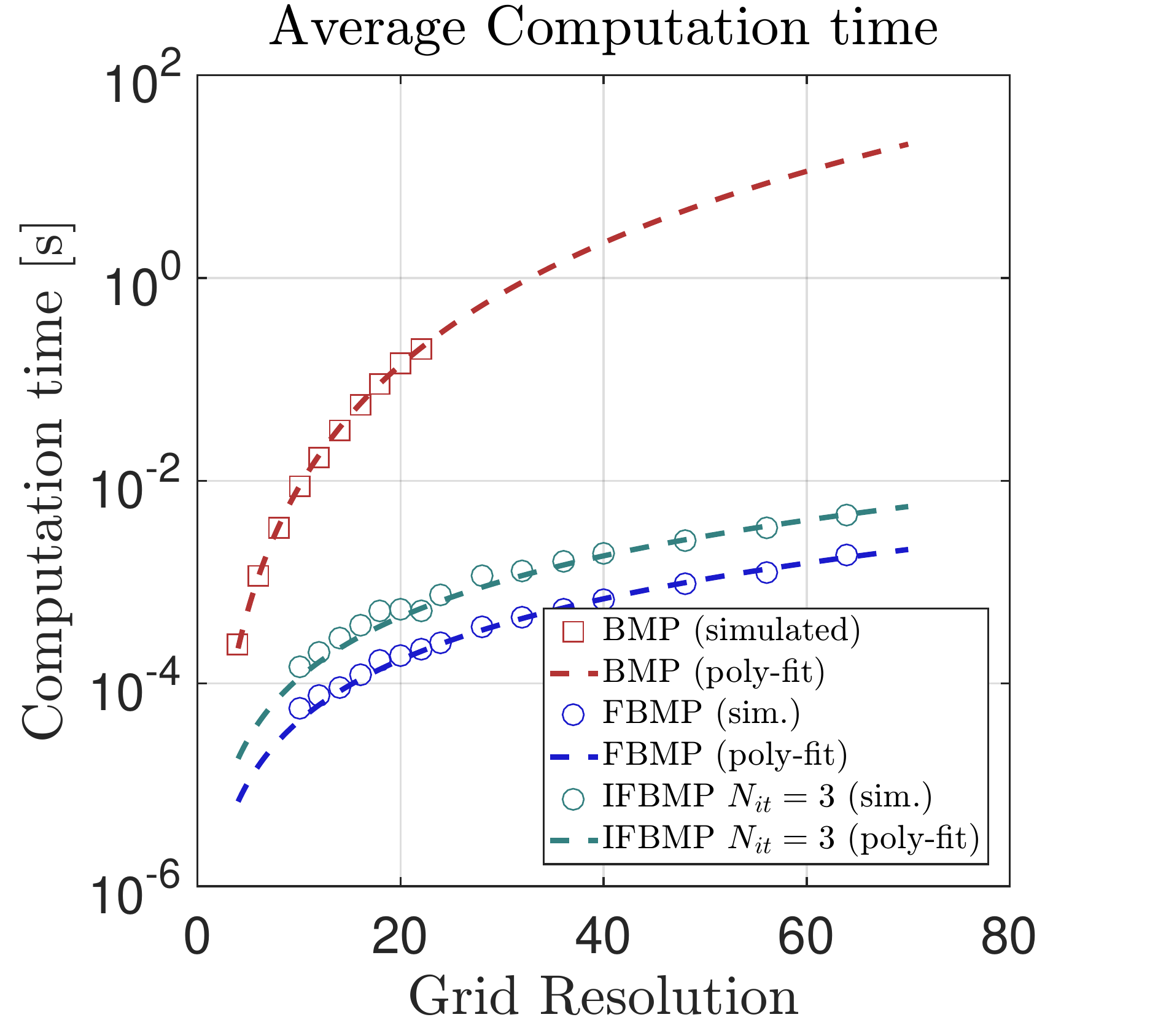}\\[-\hlabel]%
(a)\vspace{\hlabel}}\hspace{-4mm}
%\raisebox{-.5mm}{\includegraphics[width=0.275\textwidth]{img/comp_time_gr_a.pdf}}\hspace{-4mm}
%%%%%%%%% Fig 1.(b)
\parbox[b]{0.285\textwidth}{%
\noindent\includegraphics[width=0.285\textwidth]{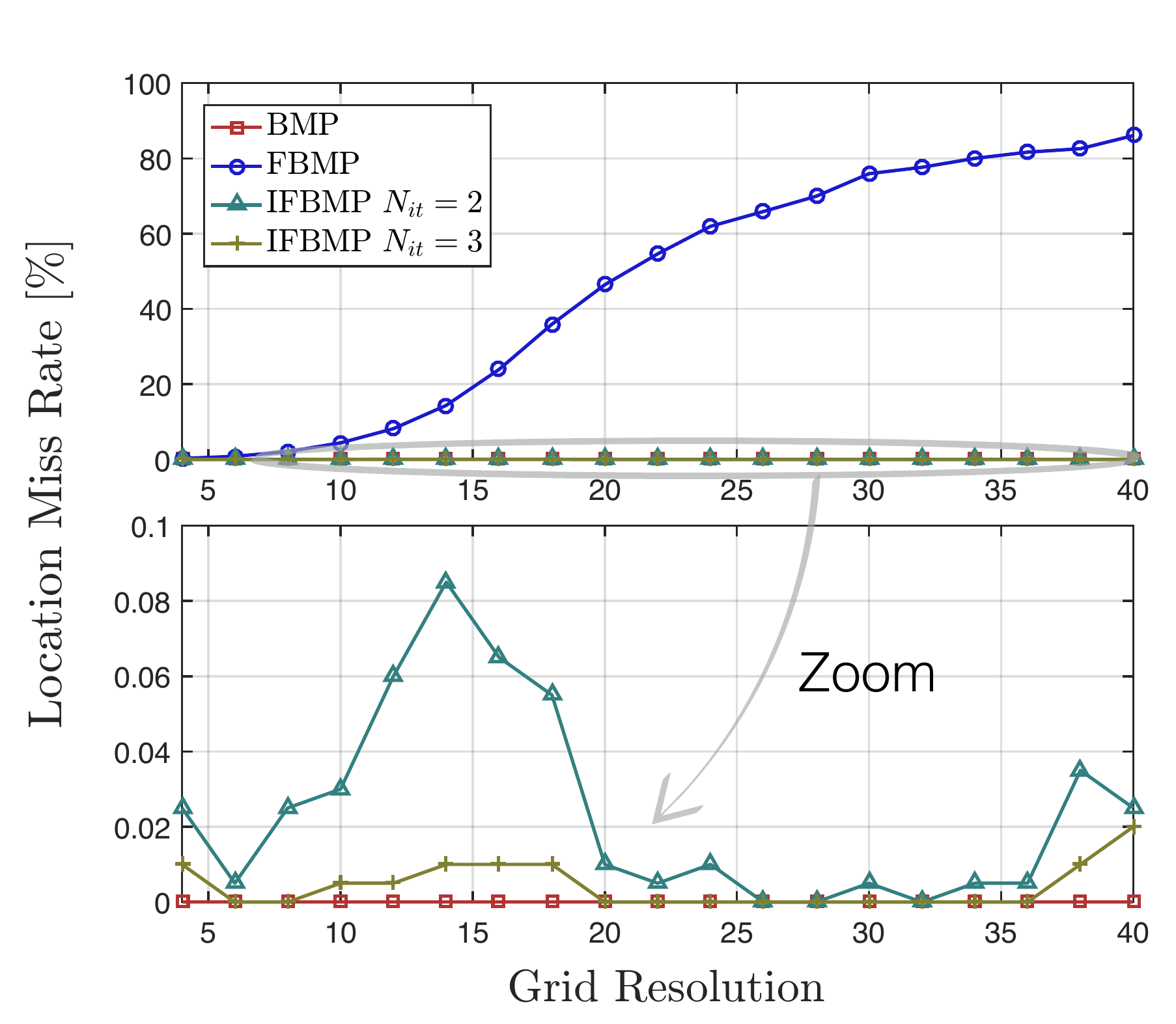}\\[-\hlabel]%
(b)\vspace{\hlabel}}
%%%%%%%%% Fig 1.(c)
\parbox[b]{0.29\textwidth}{%
\noindent\includegraphics[width=0.29\textwidth]{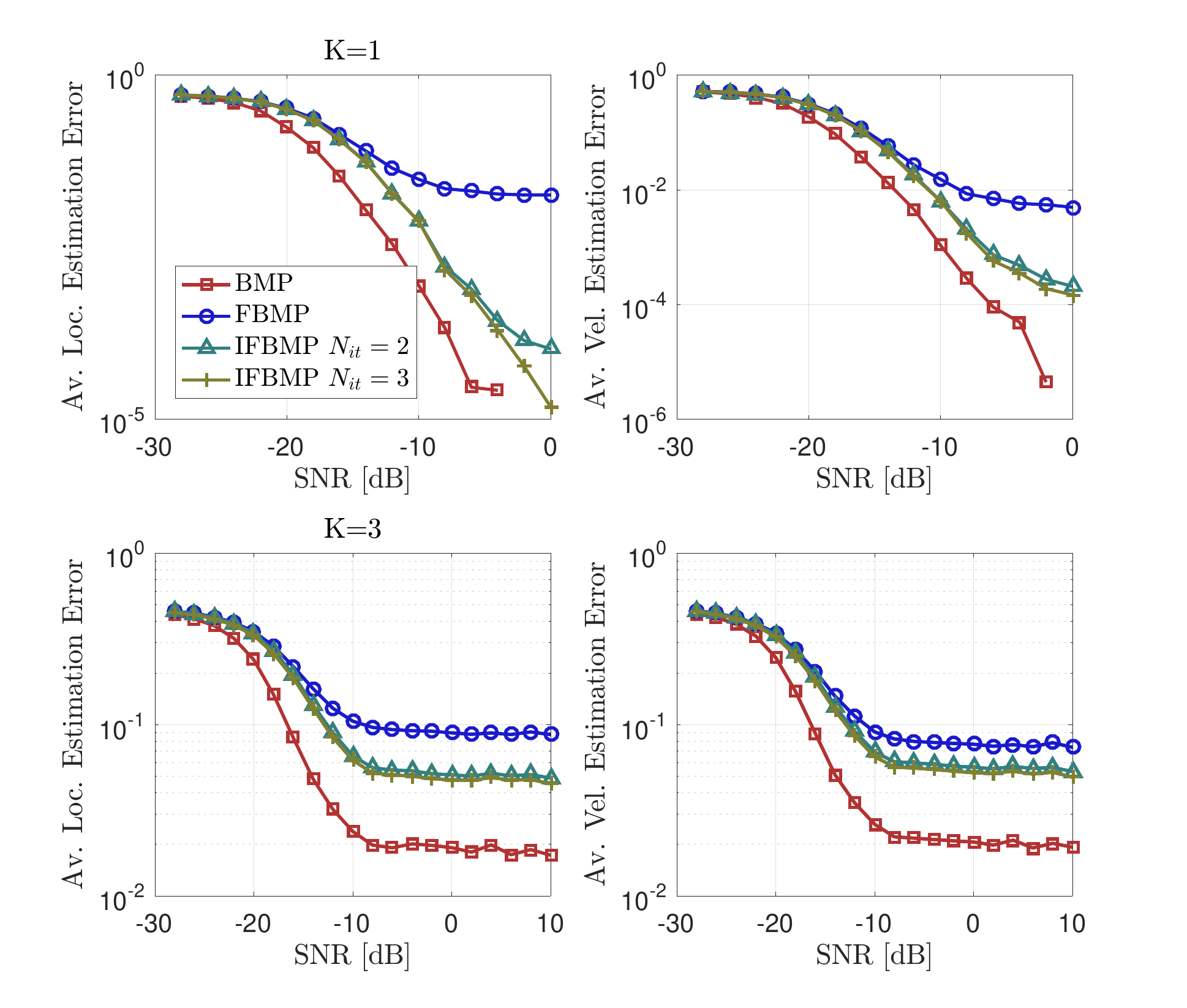}\\[-\hlabel]%
(c)\vspace{\hlabel}}
%%%%%%%%% Fig 1.(d)
\parbox[b]{0.15\textwidth}{%
\noindent\includegraphics[width=0.147\textwidth]{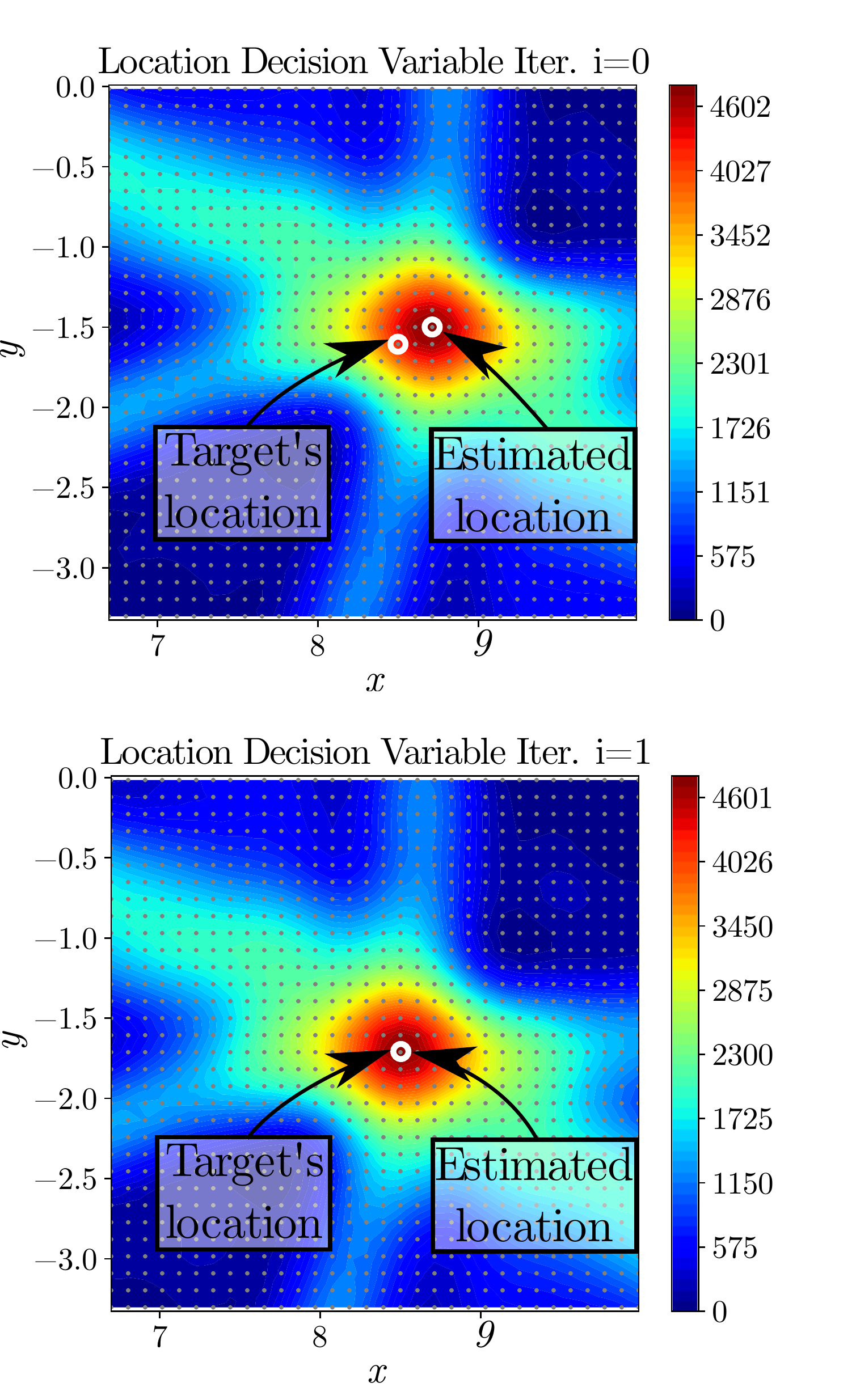}\\[-\hlabel]%
\phantom{a}\hspace{-3mm}(d)\vspace{\hlabel}}\vspace{-5mm}
%%%%%%%%% Caption
\caption{(a)~Average computation time of BMP and (I)FMP in function of the grid resolution. The polynomial fit of the BMP comp. time is $8.68\,10^{-7} \xi^4$ s. The poly-fit of the FBMP comp. time is $4.28\,10^{-7} \xi^2$ s. The poly-fit of the IFBMP comp. time is $1.12\,10^{-6} \xi^2$ s. (b)~Location Miss Rate vs the grid resolution for a single target, for an increasing number of iterative corrections. (c)~Evolution of the average estimation error (right: location, left: velocity) with respect to the SNR with $K=1$ (top), $K=3$ (bottom). (d)~Example of mapping of the decision variable for the location estimation used with IFBMP, for $K=1$ when $i=0$ (top) and $i=1$ (bottom).\vspace{-6mm} \label{fig_Scene_Time}}
\end{figure*}

%    \ts \norm{\hat{\bs{v}}_{(i)} - \bs{v}}_2 < \norm{\hat{\bs{v}}_{(i-1)} - \bs{v}}_2 \Rightarrow \norm{\hat{\bs{x}}_{(i+1)} - \bs{x}}_2 < \norm{\hat{\bs{x}}_{(i)} - \bs{x}}_2.
    %\label{eq_alg_convergence1}
%\end{equation*}
%Finally, the property
%\begin{equation*}
%    \ts \norm{\hat{\bs{x}}_{(i+1)} - \bs{v}}_2 < \norm{\hat{\bs{x}}_{(i)} - \bs{v}}_2 \Rightarrow \norm{\hat{\bs{v}}_{(i+1)} - \bs{x}}_2 < \norm{\hat{\bs{v}}_{(i)} - \bs{x}}_2.
    %\label{eq_alg_convergence2}
%\end{equation*}

%is strongly dependant of the nodes' locations, yet holds for widely spread node locations defining a multistatic system viewing the scene from very different angles, as shown by the extensive simulations of the next section. The two equations above describe the convergence of IFBMP in noiseless condition when $K=1$.
%~\eqref{eq_alg_convergence1} and~\eqref{eq_alg_convergence2}

\section{Monte-Carlo evaluation}

%\GM{A peut-être insérer qq part: We actually do the same thing with $\ox$, with the more general formulation: $|d_{b,max}-d_{b,min}| \leq M\frac{c}{B}$, which is a generalization of the Shannon-Nyquist with the reference frequency band non-centered in $f=0$. For velocity we naturally center it in $f=0$, hence in $\bs{v}=\bs{0}$ because there is no reason to look in one direction of velocity than another.}

\label{sec_simu}
%The three algorithms BMP, FBMP and IFBMP share the three last steps of Alg.~\ref{alg_BMP_applied} but differ in the computation of $\left\{n^*, {\dot n}^*\right\}$ at each iteration. Because BMP uses the complete exact model, we expect it to exhibit the best performance with accurate reconstruction in noiseless condition if $K=1$. Both FBMP and IFBMP provide approximate estimations of $\left\{n^*, {\dot n}^*\right\}$ but shorten the computing time. The first of the simulation is performed with $K=1$ in noiseless condition to highlight the estimation errors of FBMP caused by the model mismatch and the correction IFBMP provides. 
The three algorithms BMP, FBMP and IFBMP share the three last steps of Alg.~\ref{alg_BMP_applied} but differ in the computation of $\left\{n^*, {\dot n}^*\right\}$ at each iteration. Because BMP uses the complete exact model, we expect it to exhibit the best performance with accurate reconstruction in noiseless condition if $K=1$. Both FBMP and IFBMP provide approximate estimations of $\left\{n^*, {\dot n}^*\right\}$ but they shorten the computing time. The first of the simulation highlights the estimation errors of FBMP caused by the model mismatch and the correction brought by IFBMP. They are therefore performed with $K=1$ in noiseless conditions. We also assumed $\alpha^q_k\IID \cn{1}$, $q\in [Q], k\in [K]$.

%We simulated the system represented in Fig.~\ref{fig_Scene_Time}(left)
\paragraph*{a) Simulated radar System:} The location of the antennas and the grids' properties are parameters that can be optimized according to different criterion~\cite{ivashko2016, leusa2017} but are arbitrarily chosen in this work. We simulated a multistatic FMCW radar system composed of 2 TX nodes located in $\bs{x}^{(1)}_{TX} = (0, -2.5)$, $\bs{x}^{(2)}_{TX} = (7.5, -10)$ and 2 RX nodes located in $\bs{x}^{(1)}_{RX} = (0, 2.5)$, $\bs{x}^{(2)}_{RX} = (12.5, -10)$, constituting $Q=4$ bistatic pairs. The bandwidth of the signals is $B=250MHz$ and the lower carrier frequency is $f_0=24GHz$ (K-band). Each bistatic pair provides a sampled signal with $M_r = M_s=16$ at a sampling rate $\frac{1}{T_s}=50kHz$. The grids, arbitrarily chosen squared and uniformly sampled, are defined unambiguously according to Shannon-Nyquist: $\ov$ defines a square centered in $\bs v=\bs 0$ of length $L_v = \frac{\cel}{2\sqrt{2}f_0T}$, which matches \eqref{eq_alg_vbounded}; and $\ox$ defines a square of length $L_x = M_s\frac{\cel}{2\sqrt{2}B}$ with its bottom left corner in $\bs x = (5, -5)$, which matches the more general formulation of the Shannon-Nyquist sampling theorem with the frequency band non-centered in $0$, fulfilled by the sufficient condition $|\max_{\bs x\in\ox} d^q_{b}(\bs x)- \min_{\bs x\in\ox} d^q_{b}(\bs x)| \leq M_s\frac{c}{B}$ for all $q\in[Q]$. The default sampling of these square grids are respectively $M_s\times M_s$ and $M_r \times M_r$, thus $N_x=M_s^2$ and $N_v=M_r^2$.

\paragraph*{b) Algorithms Evaluation:} To evaluate estimation errors caused by the mismatch and validate the lower computational cost promised by our algorithms, we swept the value of the \emph{grid resolution} $\xi = \sqrt{N_x} = \sqrt{N_v}$. For each grid resolution, we simulated 20,000 times a target randomly selected \emph{on the current grid}, and estimated its parameters using BMP and (I)FBMP. Fig.~\ref{fig_Scene_Time}(a) confirms that the average computation times to compute $\left\{n^*, {\dot n}^*\right\}$ fit the complexities $O(\xi^2)$ and $O(\xi^4)$ respectively expected for, (I)FBMP and BMP. Fig.~\ref{fig_Scene_Time}(b) confirms that, in noiseless condition BMP makes no estimation error, while the miss rate of FBMP --- the rate of wrong cell selection in the grids --- grows toward almost 100\% when the grid gets very dense because of the shift described in Thm.~\ref{th_1}, which is more often accounted for with dense grids. The average location estimation error grows from $0\%$ to $2.5\%$ when $\xi\leq 16$, then remains constant when $\xi>16$. In short, low-density grids lead to few but large errors while high-density grids lead to small but systematic errors. The zoomed figure (Fig.~\ref{fig_Scene_Time}(b)) shows that the location errors of FBMP are almost completely compensated for within a very small number of iterations with IFBMP. An example of detection of a target in $\bs x = (8.5, -1.7)$ and $\bs v = (6,6)$ using two iterations of IFBMP is provided in Fig.~\ref{fig_Scene_Time}(d). The robustness against noise is evaluated from an operating point set to $\xi=16$. Figure~\ref{fig_Scene_Time}(c) shows the evolution of the average location and velocity estimation errors (LEE and VEE, respectively), defined from the parameters estimates $\hat{\bs x}_k$ and $\hat{\bs v}_k$, $k\in[K]$ as
$$
\ts {\rm LEE} \is \frac{1}{K}\sum_{k\in\set{K}} \frac{\norm{\hat{\bs{x}}_{k}-\bs{x}_{k}}_2}{L_x}, \ {\rm VEE} \is \frac{1}{K}\sum_{k\in\set{K}} \frac{\norm{\hat{\bs{v}}_{k}-\bs{v}_{k}}_2}{L_v}, 
$$
in function of the signal to noise ratio (SNR), defined from $\sigma_e^2$, the common noise power for all bistatic pair, as
\begin{equation}
    \ts \text{SNR} = \frac{1}{K}\frac{\mathbb{E}\left(\sum_{q\in[Q]} \norm{\bs{D}^q\bs{s}^q}_2^2\right)}{\mathbb{E}\left(\sum_{q\in[Q]} \norm{\bs{e}^q}_2^2\right)} = \frac{1}{\sigma_e^2}.
\end{equation}

IFBMP exhibits a lower robustness against noise than BMP as a result of the non-coherent summation performed in~\eqref{eq_alg_fbmpnx}. Moreover, when $K>1$, IFBMP goes toward a higher error value than BMP when the SNR goes to infinity. The main effect occurring here is the detection of ghost targets~\cite{ohagan2018}.%, which can be strongly counteracted by hybrid multistatic MIMO systems.
%In a nutshell, these simulations confirms that both BMP and IFBMP, that proposed as an alternative to BMP, provide perfect target parameters estimations in noiseless condition.

In a nutshell, our simulations confirm that IFBMP, that we proposed as an alternative to BMP, provides the same quality of target parameters estimations as BMP in noiseless condition if $K=1$ while being faster and practical for dense grids. Yet, its smaller complexity comes at the price of a lower robustness against noise.

%\paragraph*{c) Algorithms Comparison:} BMP exhibits the best performance with perfect target parameters estimations in noiseless condition and a best robustness against noise. FBMP is faster and practical for dense grids but is affected by a drift in the estimations of location. In noiseless condition, IFBMP is able to perfectly correct those errors. Still, the smaller computational cost of (I)FBMP comes at the price of a lower robustness against noise. 

%However, the computational cost is not scalable for dense grids reaching the resolution radars offer. Therefore, the algorithm is not practical for real applications.

\section{Conclusions and perspectives}
\label{sec_concl}
%and offered an analysis of the impact of the simplified modelling
While algorithms for sparse signals reconstruction have gained in popularity in target detection with multistatic radars, the proposed adaptations such as the Block Matching Pursuit are computationally not scalable. We have shown how an explicit formulation of the dictionaries obtained from FMCW chirp-modulated signals enable simplifications and methods for dramatically reducing this complexity with limited losses of precision, as shown by Thm~\ref{th_1}. The comparison of BMP with (I)FBMP we introduced confirmed the lower computational cost of our methods and showed that the precision losses are compensated while keeping identical computational complexity when using IFBMP. Therefore, this work is expected to allow affordable use of sparse reconstruction for FMCW multistatic radars and further lead to the design of multistatic FMCW compressive radars. Yet, the overall geometry of both the radar system and the grids provide degrees of freedom to be optimized with the help of the derivation of theoretical bounds such as Cramer-Rao bounds or Ziv-Zakai bounds.

%TX and RX nodes' locations and the geometry of the grids are degrees of freedom to be optimized with the help of the derivation of theoretical bounds such as Cramer-Rao bounds or Ziv-Zakai bounds, as well as the study of the self-coherence of the dictionaries. 
%Such work is related to the optimization of the radars' location and to the derivation of theoretical bounds such as Cramer-Rao bounds or Ziv-Zakai bounds, as well as the self-coherence of the dictionaries. 

\appendix
\section{Proof of Theorem~\ref{th_1}}
Considering~\eqref{eq_mod_psidef}, the only factor of $\psi^q_{\bs{x},\bs{v}}[m_s]$ that depends on $m_s$ is
$\rho^q_{\bs x, \bs v}[m_s] := \exp(-j \frac{2\pi}{\cel} (\frac{B}{M_s} r^q(\bs x) + f_0 T_s v^q(\bs x, \bs v)) m_s)$.
We must thus prove that 
$|\rho^q_{\bs x, \bs v}[m_s] - \rho^q_{\tilde{\bs x}, \bs 0}[m_s]| =  O(\frac{1}{\lambda})$.

We first note that $\|\bs{x}_{\rm t}^q- \tilde{\bs x}\|_2 = \|\bs{x}_{\rm t}^q- \bs x\|_2 \sqrt{1 + \kappa}$ with
$\ts \kappa := 2\gamma \sprod{\bs{x}_{\rm t}^q-\bs{x}}{\bs{v}} \norm{\bs{x}_{\rm t}^q-\bs{x}}^{-2}_2 + \gamma^2 \norm{\bs{v}}^2_2\norm{\bs{x}_{\rm t}^q-\bs{x}}^{-2}_2$.
Using~\eqref{eq_alg_vbounded}, $T = M_sT_s$, and \eqref{eq_alg_radoutside}, we find $\gamma\norm{\bs{v}}_2 \leq \frac{c}{4B} <  \frac{1}{\lambda} \norm{\bs{x}_{\rm t}^q-\bs{x}}_2$. 

Therefore, we show easily that $\kappa \geq \lambda^{-2} - 2\lambda^{-1} \geq -1$ and $\kappa \leq 3 \lambda^{-1}$, and since $1+\frac{s}{2}(1 - s)\leq \sqrt{1+s}\leq 1+\frac{s}{2}$ for all $s \geq -1$ and $\gamma^2 \frac{\norm{\bs{v}}^2_2}{\norm{\bs{x}_{\rm t}^q-\bs{x}}_2} \leq \frac{\gamma \norm{\bs{v}}_2}{\lambda}$, $\|\bs{x}_{\rm t}^q- \tilde{\bs x}\|_2$ is smaller than\vspace{-2mm}
$$
\ts \|\bs{x}_{\rm t}^q- \bs x\|_2(1 +\frac{\kappa}{2}) \leq \|\bs{x}_{\rm t}^q- \bs x\|_2 + \gamma \frac{\sprod{\bs{x}_{\rm t}^q-\bs{x}}{\bs{v}}}{\norm{\bs{x}_{\rm t}^q-\bs{x}}_2} + \frac{\gamma \norm{\bs{v}}_2}{2\lambda},\vspace{-2mm}
$$\vspace{-2mm}
and larger than%\vspace{-0.5mm}
\begin{multline*}
\ts \|\bs{x}_{\rm t}^q- \bs x\|_2(1 +\frac{\kappa}{2}(1 - \kappa)) \geq \|\bs{x}_{\rm t}^q- \bs x\|_2(1 +\frac{\kappa}{2}(1 - \frac{3}{\lambda})),\\
\ts \geq \|\bs{x}_{\rm t}^q- \bs x\|_2 + \gamma \frac{\sprod{\bs{x}_{\rm t}^q-\bs{x}}{\bs{v}}}{\norm{\bs{x}_{\rm t}^q-\bs{x}}_2}\,(1 - \frac{3}{\lambda})\\
\ts \geq 
\|\bs{x}_{\rm t}^q- \bs x\|_2 + \gamma \frac{\sprod{\bs{x}_{\rm t}^q-\bs{x}}{\bs{v}}}{\norm{\bs{x}_{\rm t}^q-\bs{x}}_2} - \frac{3 \gamma \|\bs v\|_2}{\lambda}.\vspace{-2mm}
\end{multline*}
Using the definition \eqref{eq_mor_bispeeddef}, and adding to the bounds above similar bounds on $\|\bs{x}_{\rm s}^q- \tilde{\bs x}\|_2$, we find from \eqref{eq_alg_vbounded} \vspace{-1mm}
\begin{equation*}
    \ts \big|\,r^q(\tilde{\bs x}) - \big(r^q(\bs{x})+\gamma v^q(\bs{x},\bs{v})\big)\,\big|\ =\  O(\frac{\gamma \|\bs v\|_2}{\lambda}) = O(\frac{\cel}{\lambda B}).
\vspace{-1mm}
\end{equation*}
Finally, since $\big|\rho^q_{\bs x, \bs v} - \rho^q_{\tilde{\bs x}, \bs 0}| \leq |\angle\rho^q_{\bs x, \bs v} - \angle\rho^q_{\tilde{\bs x}, \bs 0}|$, we get $\big|\rho^q_{\bs x, \bs v}[m_s] - \rho^q_{\tilde{\bs x}, \bs 0}[m_s]\big| \leq \frac{2\pi m_s B}{\cel M_s}|r^q(\tilde{\bs x}) - \big(r^q(\bs{x})+\gamma v^q(\bs{x},\bs{v}))| = O(\frac{\pi}{\lambda})$, by using \eqref{eq_alg_vbounded} and $\frac{M_s}{B} f_0 T_s = \gamma$. This concludes the proof.

\bibliographystyle{unsrt}
\bibliography{main.bib}

\end{document}